# Fluctuations in growth rates determine the generation time and size distributions of *E. coli* cells.

Mats Wallden[‡], David Fange[‡], Özden Baltekin, Johan Elf[†]

Science for Life Laboratory, Uppsala University, Sweden

[‡] Equal Contribution [†]Johan.Elf@icm.uu.se

**Isogenic *Escherichia coli* growing exponentially in a constant environment display large variation in growth-rates, division-sizes and generation-times. It is unclear how these seemingly random cell cycles can be reconciled with the precise regulation required under conditions where the generation time is shorter than the time to replicate the genome. Here we use single molecule microscopy to map the location of the replication machinery to the division cycle of individual cells. We find that the cell-to-cell variation in growth rate is sufficient to explain the corresponding variation in cell size and division timing assuming a simple mechanistic model. In the model, initiation of chromosome replication is triggered at a fixed volume per origin region, and associated with each initiation event is a division event at a growth rate dependent time later. The result implies that cell division in *E. coli* has no other regulation beyond what is needed to initiate DNA replication at the right volume.**

Bacterial cells growing at the same rate have largely the same composition of their major components independent of growth media(*1*)(*2*). This composition is determined by the bacterial growth laws (*3*) and implemented by transcriptional feedback systems sensitive to imbalances in metabolic fluxes (*4*). Balanced exponential growth further requires that both DNA replication and cell division keep up with the increasing biomass, despite that for rapidly growing *E. coli* it takes longer to replicate a chromosome than to double the biomass. A solution to this problem was presented by Cooper and Helmstetter(*5*), who proposed that cell division occurs a constant time after initiation of DNA replication (Fig. 1A). For rapidly growing cells this implies that new replication rounds are started before the previous round has finished, since replication has to start on average once per generation in order to get a stable cell cycle. Such stable cell cycles are demonstrated by a deterministic simulation in Fig. 1B. Recent measurements of the generation times and divisions size for individual cells suggest that division is not triggered by replication, but when the cell has grown by a certain volume that is independent of the birth volume (*6*). This seems to imply that the replication and division cycles are largely independent. We now ask how individual cells make sure that chromosomes are replicated and segregated in time, especially considering the large cell-to-cell variation observed in generation time and cell size(*7, 8*).

We have directly studied the coordination of the replication and cell division cycles in *E. coli* by imaging fluorescently labelled epsilon subunits of DNA Pol III, named DnaQ (*9*), in exponentially growing cells. Due to the low diffusivity of the Pol IIIs engaged in replication, individual replisomes can be localized using single molecule fluorescence imaging (Fig 2A). The labelling does not influence growth (Fig S1 A,B) or replication initiation (Fig S1C). Cell size and division events were determined by time-lapse phase contrast microscopy of cells grown in a microfluidic device (Fig. 2C). The precision in division time determined by phase contrast was found to be 2 minutes based on a comparison to a fluorescent segmentation marker (Fig S2). The growth rates determined by phase contrast and fluorescence microscopy for individual cells were the same (Fig S3D). Individual cells grow exponentially, independently of the position within the device and unperturbed by imaging laser exposure (Fig S3A-C). In agreement with previous reports (*6, 7, 10*), we observed large cell-to-cell variations in division size, generation time and growth rates (Fig 2D,E,F). By combining the phase contrast imaging of growth and division with fluorescence imaging of DnaQ localization we can now map the replication cycle to the division cycle for individual cells growing at different rates.

Figure 3A-C shows the localization of replisomes along the long axis of the cells as a function of the cell volume, for three different growth conditions(*11*). New rounds of replication are observed to start at defined cell sizes (Fig. 3A-C, white line). If the cells instead are aligned by the time from division, the time of appearance of new replisomes is unclear (Fig S4). This suggests that the control of the replication cycle is related to size rather than the time from division (*12, 13*). The origin of replication locus co-localizes with the replisomes at replication initiation, as shown by a fluorescent MalI transcription factor bound at the *oriC*-proximal *bgLG* locus (Fig S5).

Furthermore, the number of ongoing replication cycles were validated by replication runout experiments (*14*) adapted to the microfluidic environment (SI). For slow, intermediate and fast growth, the number of origins are typically 1 or 2, 2 or 4 and 4 or 8, respectively (Fig S6). The observed initiation volumes (Fig. 3A-C, white lines) divided by the corresponding number of origins is relatively invariant for the different growth conditions (~0.9μm$^3$). This is particularly clear for slow growth, where a significant fraction of the cells initiate one round of replication at 0.9μm$^3$ and another at 1.8μm$^3$ (Fig 3A). Initiation timing has to be exceptionally sensitive to changes in the chromosome to volume ratio in order to trigger replication only in the experimentally observed range. In fact, we found that no previously described molecular mechanism(*15*) gives the required sensitivity and robustness. In the supplementary material we therefore show how sufficiently sensitive initiation can be implemented by a zero order switch(*16*) in the DnaA-ATP to DnaA-ADP ratio.

The mapping of replication to the division cycle also allows us to make a direct determination how much time cells are given to replicate and segregate their chromosomes before division, i.e. the C and D periods, for the different growth conditions (Fig 3D, E). We find that the average C+D-period, $\tau$, is relatively constant for the two fastest growth conditions but much longer at slow growth (Fig 3F). Previous replication run-out experiments(*17*) in bulk suggest a functional form for the growth-rate dependence as $\tau=\alpha\mu^{-\beta}+\gamma$. This form can, for example, be obtained if chromosome replication, chromosome segregation and septum formation at fast growth takes a minimal time, $\gamma$, due to the multiple sequential steps involved; and that these three processes on the other hand are limited by protein synthesis and metabolism at slow growth, where $\tau$ increases by $\beta$ percent for each percent decrease in growth rate. To test if the $\tau$ obtained for the population averages applies also for individual cells we stratified the data based on growth rate and determined $\tau$ for each category (Fig 3F, circles). Using this strategy we show that there is growth-rate dependence of the C+D period, $\tau$, also at the single-cell level.

Given the success in the population based models for how growth-rates dictate average properties of cells(*18*), one may ask if an individual cell's growth rate also dictates its division timing. We do this by recasting the Cooper and Helmstetter model

(Fig.1A) in a stochastic single cell setting (Fig. S9) where we can test if the cell-to-cell variation in growth rate alone can predict the observed generation time and division size distributions. The assumptions in our model are: 1. The individual cell's growth rate is sampled from the experimentally observed distribution (Fig 2 F ). 2. Replication is initiated at a fixed volume per origin ratio (fig 3A-C). 3. The cell is divided at a growth rate dependent time after initiation (Fig 3F). In addition, we account for experimentally observed mother-daughter and sibling correlations in growth rate (Fig S7CD) and also include 15% inaccuracy (two standard deviations) in the volume of replication initiation, which matches the inaccuracy allowed to initiate multiple origins within the sequestration period (*19*). Given these assumptions we can predict how variation in growth rate at the single cell level propagates to predicted variation in cell size and generation time (Figs.4, S7).

Figure 4 shows the predicted distributions and covariation for a number of cell cycle related parameters assuming that cell-to-cell variability originates in growth rate. There is good agreement with the corresponding observations, suggesting that the model captures the most important aspect of cell cycle control and its variability. We therefore use the model to explain a number of prominent features of cell cycle control. For example, with regard to the covariation of division sizes, $V_D$, and generation times, T, (Fig. 4A) the average volume at division is accurately predicted (dotted black line) by $V_D=V_I e^{\mu\tau}$, where $V_I$ is the fixed initiation volume per origin $\tau=\alpha\mu^{-\beta}+\gamma$ (Fig 3F) and the average growth rate is $\mu=\ln(2)/T$. Secondly, the birth size has strong negative correlation with the generation time for rapidly growing cells (Fig. 4A solid black contours). This is due to the fact that large new-born cells originated from rapidly growing mothers that reached their initiation volume earlier and therefore spent more of the C+D-period in their own cell cycle than that of their daughter. Lastly, the division sizes are uncorrelated with generation time for slowly growing cells, a consequence of the C+D period's strong dependence on the growth rate (Fig 3F). This implies that slow cells have more time to grow from the constant replication initiation volume to cell division (Fig 4D). For example, in the limit of slow growth $\tau= \alpha\mu^{-\beta} \approx \alpha/\mu$, which results in $V_D=V_I e^{\alpha}$, i.e. a constant division size. This behaviour appears as a "sizer", although the main reason is not an explicit size sensor but rather that the C+D period is growth rate dependent. Furthermore, under conditions with overlapping replication cycles, the individual cells growth rate is uncorrelated with the time to its division, as this event is triggered in the mother or grandmother generation (Fig. 4E). In effect this will give a volume expansion per generation that is uncorrelated to the birth volume (Fig 4C), which in previous descriptions has been referred to as an "adder" (*6, 8*). Our replication-triggered division model with growth rate dependent C+D periods thus reconciles Cooper-Helmstetter's overlapping replication cycles with recent single cell observations and also correctly predicts deviation from the "adder" model at slow growth (Fig 4C).

We have shown that cell-to-cell variation in size and generation time only depends on the difference in growth rate between individuals. Coordination between replication and division is, however, assured by initiating DNA replication at a nearly fixed size per chromosome. The remaining major question is what determines the cell-to-cell variation in growth rate, i.e. why do all cells not grow as fast as the fastest cell for a specific growth medium? The answer is most likely related to the high cost of the control systems required to tune each cell to the maximal growth in each specific condition, which would likely result in a decrease in the average growth rate. Instead, it appears the cell uses a slightly sloppy control system that results in important components getting into a suboptimal balance. In Fig. S8 we test if the uneven inheritance of ribosomes between sisters (*20*) is correlated with growth rate and find that ribosome partitioning has no significant influence on the variation in growth rate. It appears the situation is more complicated: any deviation in a large number of balanced fluxes may lead to reduced growth, which would result in the observed high correlation between sisters' growth rate (Fig. S7D), but also a significant challenge for evolution to bypass.

## Acknowledgments

We thank Gustaf Ullman, Alexis Boucharin, Erik Marklund and Prune Leroy for valuable support in data analysis and cloning; Sajith Kecheril and Carolina Whälby for providing image analysis code; Bill Söderström for providing EC442 and Benedict Michelle for providing JJC5350. This work was supported by the European Research Council, Vetenskapsrådet, and the Knut and Alice Wallenberg Foundation.

# Figures

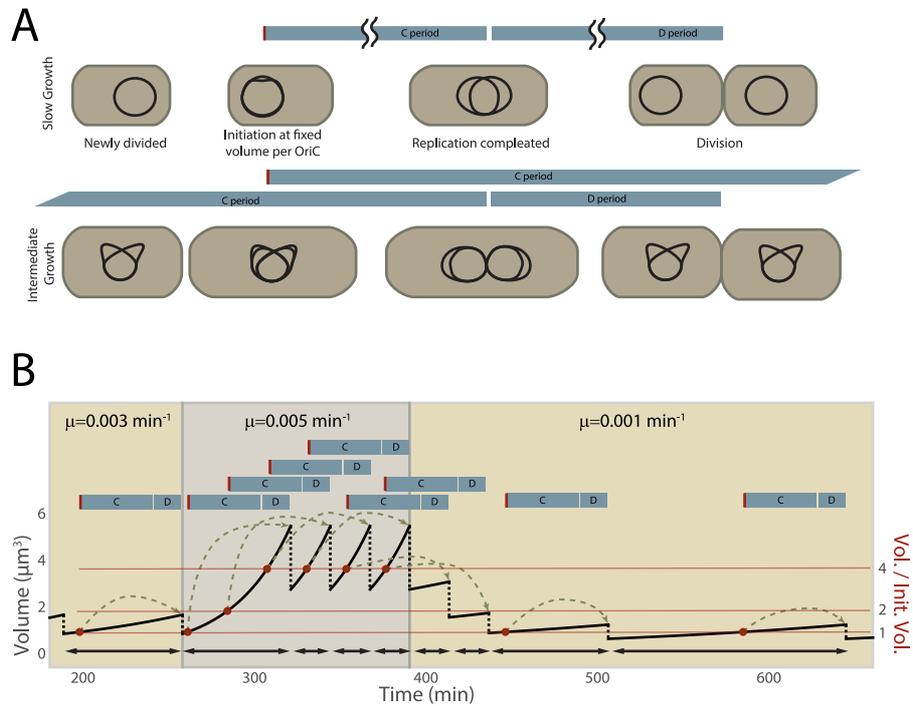

**Figure 1: Coupling of replication and division cycles in *E. coli* A.** The Cooper-Helmstetter model, where division events are scheduled a fixed time after replication initiates. Initiation once per generation leads to a requirement of overlapping replication cycles at fast growth. **B.** Simulated volume expansion and division for an idealized cell lineage going through an up-shift and a down-shift. Replication is here initiated at a fixed volume per chromosome (red circle) and the cells divide after a fixed period of time later, which includes the required time for replication (C-period), chromosome segregation and septum formation (D-period).

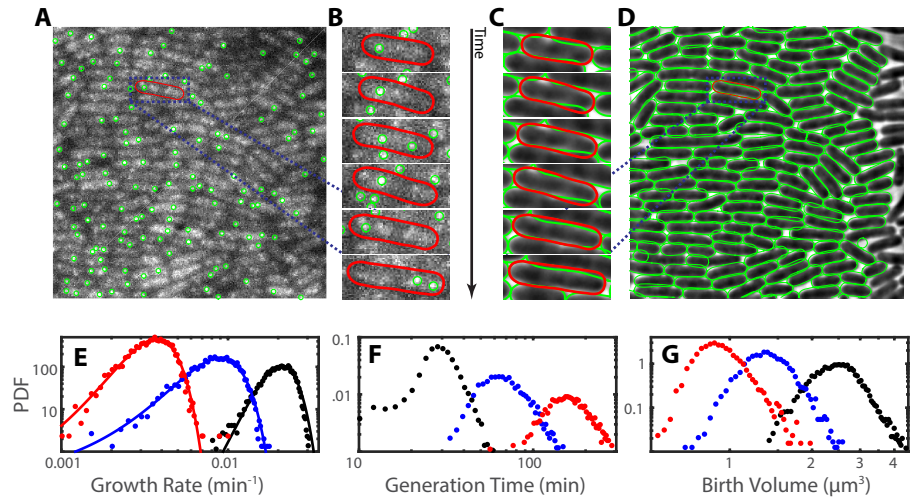

**Figure 2: Characterizing growth and chromosome replication at the single cell level. A.** Fluorescence image corresponding to D with identified DnaQ-Ypet. **B** and **C.** An individual cell tracked in time using phase contrast and fluorescence. **D.** Automatically segmented *E. coli* growing in a microfluidic device imaged using phase contrast. **E.** Distributions of single cell growth rates for fast, intermediate and slow conditions(*11*) fitted to normal distributions. The number of cells that are tracked though their entire cell cycle are 10774, 5946, 4257, respectively. **F.** Distributions of generation times corresponding to E. **G.** Distributions of birth volumes corresponding to E.

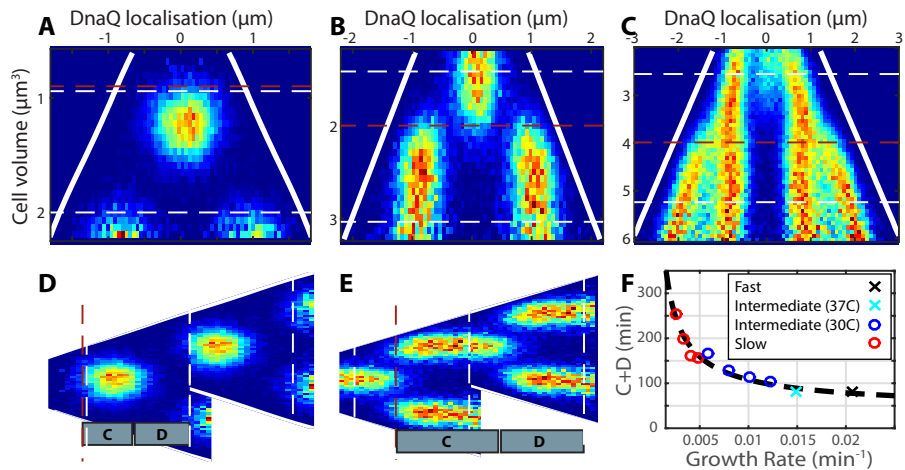

**Figure 3: Distributions of replication timing. A-C.** Distributions of localized DnaQ along the long axis of the cell (x-axis) for cells of different volumes (y-axis). A, B and C corresponds to slow, intermediate and fast growth(*11*). Replication initiation is indicated by red dashed lines. White dashed lines indicate average volumes at division. Note that the two replication-forks initiated from same origin of replication are spatially too close to be visualized as separate distributions. **D-E.** corresponds to stacking the distributions with one generation time's displacement to indicate C and D periods for slow and intermediate growth conditions respectively. **F.** The C+D periods plotted against the growth rate. The data is fitted with a power-law curve, $1.3\mu^{-0.84}+42$ min (dashed). The number of cell cycles used in each point is, in order of increasing growth rate: 593, 1427, 1464 and 595, 814, 2023, 2020, 844, 3758, and 10774.

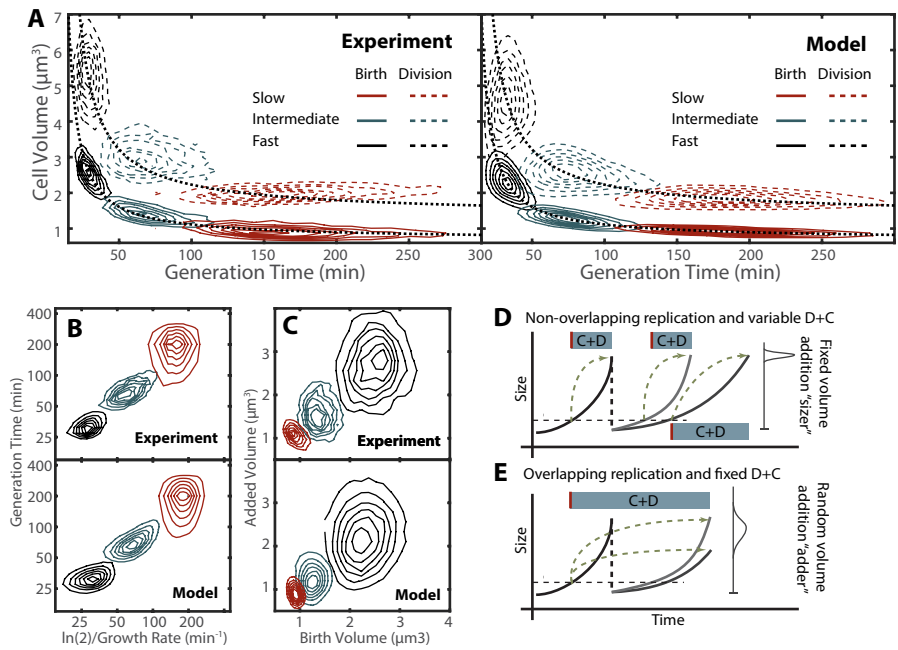

**Figure 4: The growth rate variation propagated to other variables and compared to independent experimental observations. A.** The covariation of generation time and cell size at birth (solid) and division (dashed). The dotted black lines are analytical model predictions for the average birth and division sizes for different growth rates as is described in the main text. **B.** The covariation of generation time and growth rate. **C.** The covariation of volume added during one generation and the birth volume. **D.** Example of how a model with strong growth-rate dependence of the C+D-period and non-overlapping replication cycles appear as a "sizer". **E.** Example of how a model with weak growth-rate dependence on the C+D-period appear as an "adder".

# Supplementary Figures

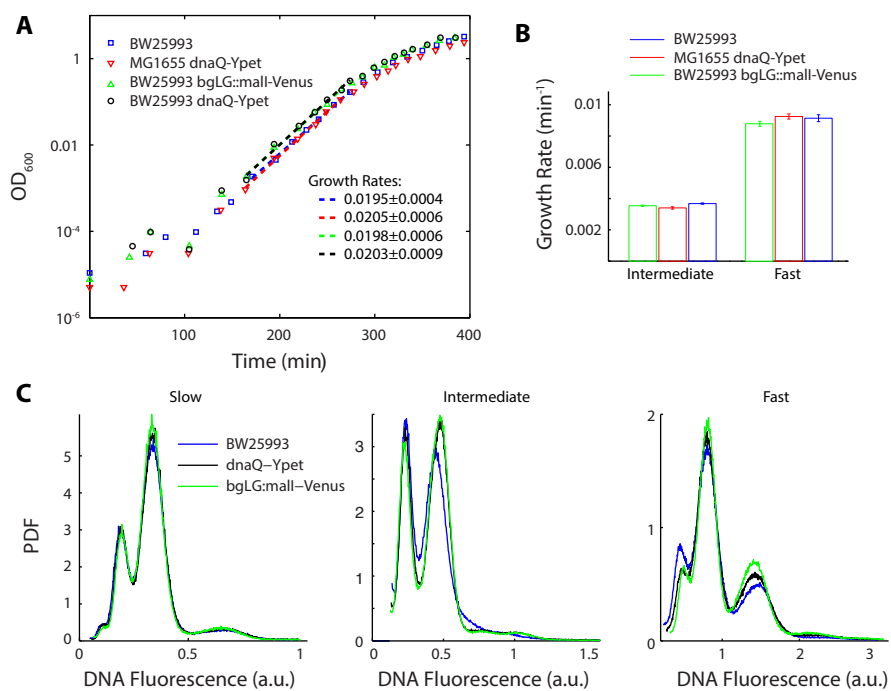

**Figure S1: Effects of labeling on growth and replication characteristics.**
(**A**) OD$_{600}$ as function of time for flask cultures under fast growth conditions (see ref. (11) of main text for definition). Exponential growth rates are estimated with linear regressions (dashed lines) and listed in the inset with the 95% confidence interval. (**B**) Exponential growth rates estimated using a commercial plate reader for intermediate (at 37°C) and fast growth conditions. Each is an average of three replicates and error bars are the standard deviation of the replicates. (**C**) Distribution of DNA fluorescence per cell for replication run-out treated flask cultures. Fluorescence values have been adjusted to align the major peaks in the distributions.

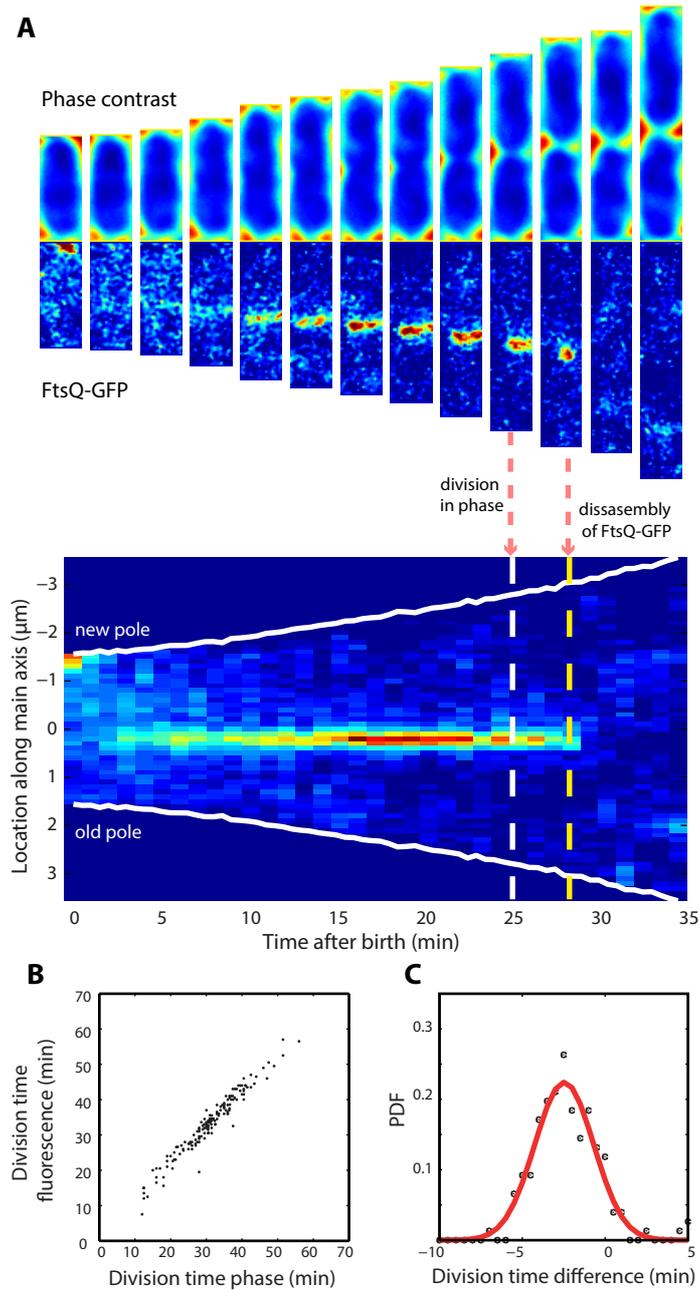

**Figure S2: Determination of division timing in phase contrast.**
(**A**) Time course of a cell expressing FtsQ-GFP imaged in phase contrast and fluorescence (*top*) and the corresponding relative average fluorescence intensity along the long axis of the cell as a function of time after birth (*bottom*). The division time based on phase contrast images (white dashed line) is compared to division time determined by FtsQ-GFP fluorescence (yellow dashed line). (**B**) The correlation between observations (n=152) of division time as determined from phase contrast and FtsQ-GFP fluorescence (r=0.97). (**C**) Distribution of the difference in division timing show in (B) and the corresponding regression to a normal distribution (red line). The estimated standard deviation of the time difference is 1.78 ± 0.18 minutes.

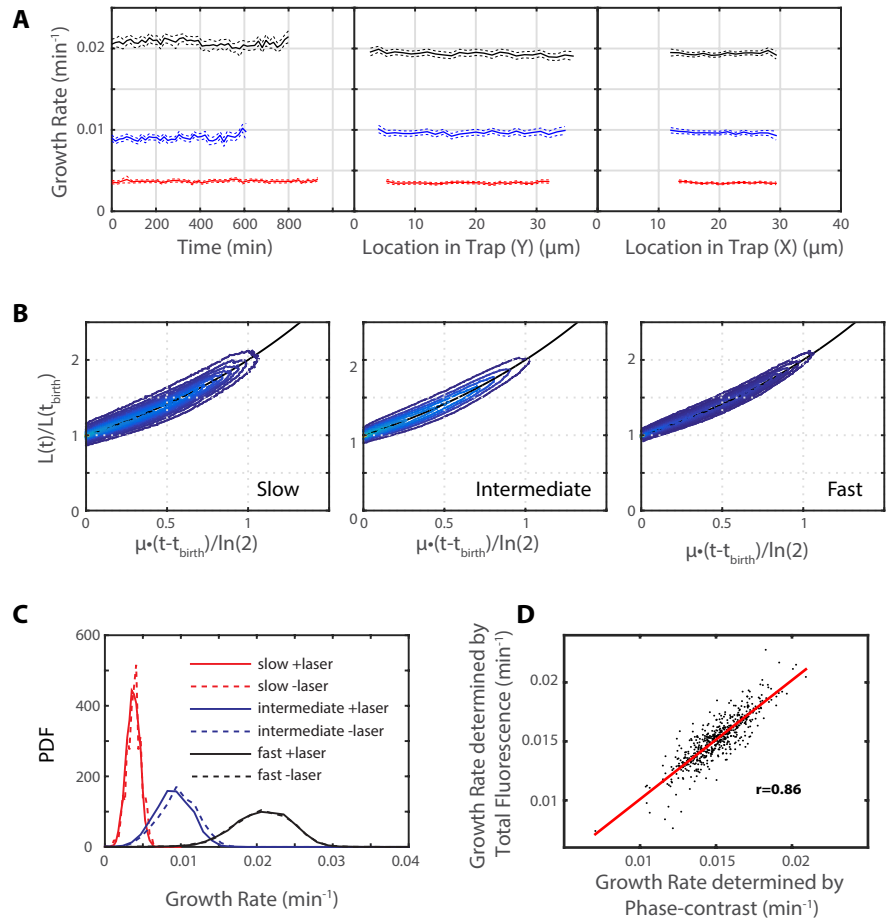

**Fig S3: Microfluidic growth characterization.**
(**A**) Average growth rates observed over the course of a time-lapsed experiment (*left*), the y-position in the traps (*center*) and the x-position (*right*) for the slow (*red*), intermediate (*blue*) and fast (*black*) growth conditions. ($n_{slow}$=4257, $n_{intermediate}$=5944, $n_{fast}$=10773). Dashed lines are the 95% confidence intervals of the average. For the time course panel (*left*) each cell cycle contributes to the average with its growth rate at the time of birth. For the position panels (*center* and *right*) each cell contribute to the average with its growth rate at the location of the average centroid position ($n_{slow}$=2832, $n_{intermediate}$=5964, $n_{fast}$=10774). The curve is truncated at the positions where the residual fraction of observations beyond it was less than 3%. (**B**) Distribution of cell lengths scaled by the birth length as function of time after division scale by $\ln(2)/\mu$ for three different growth conditions. Ideal exponential growth is shown as black solid lines. (**C**) Growth rate distributions with and without laser exposure. ($n_{slow+laser}$=4257, $n_{slow-laser}$=369, $n_{intermediate+laser}$=5946, $n_{intermediate-laser}$=1341, $n_{fast+laser}$=10774, $n_{fast-laser}$=2222). (**D**) Joint mapping of growth rates of cells determined from segmentation performed in phase contrast and fluorescence images (n=555). Linear correlation is shown as a red line (r=0.86).

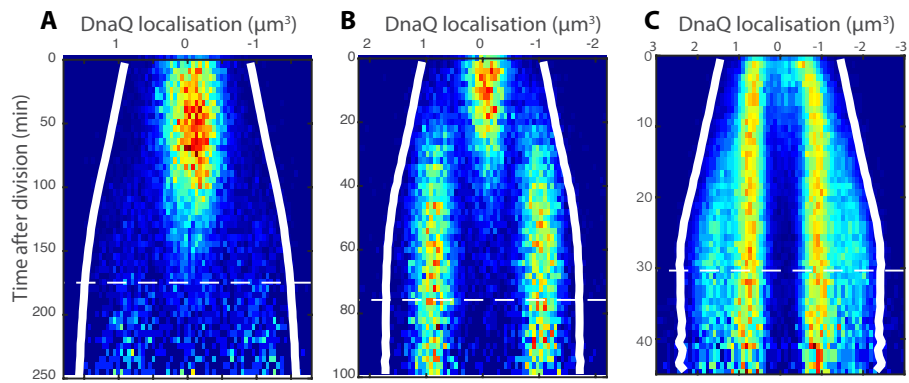

**Figure S4: Replisome localization aligned by time after birth.** As Fig. 3 A-C in the main text, but with DnaQ localization as function of time after cell division instead of cell volume. Average cell lengths (white solid lines) are not expected to grow exponentially since more cells with higher growth rates contributes in the beginning. ($n_{slow}$=4257, $n_{intermediate}$=5946, $n_{fast}$=10774).

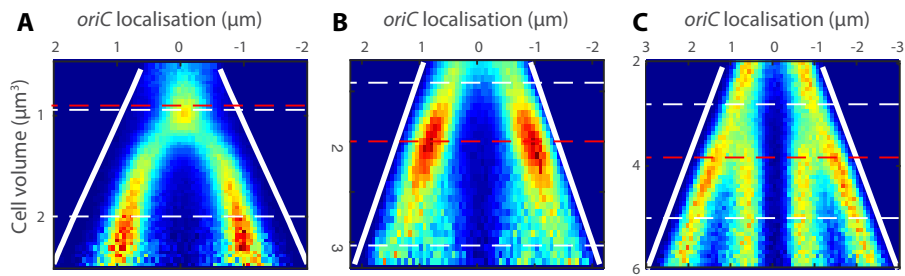

**Figure S5: Origin localization aligned by size.** As Fig 3A-C in the main text, but with localization of MalI-venus instead of DnaQ-Ypet.

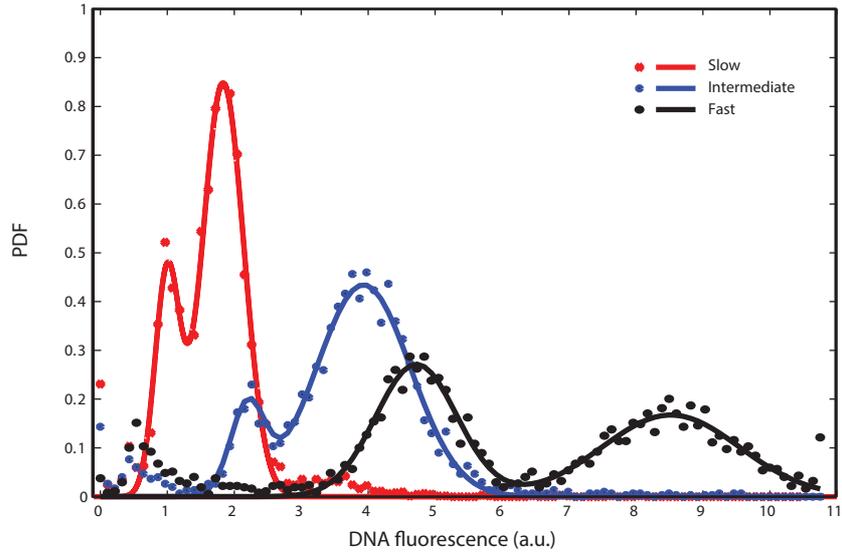

**Figure S6: Microfluidic replication run-out experiments.** Distribution of DNA-fluorescence per cell (dots) for unlabeled (BW25993) cells for three growth conditions ($n_{slow}$=5572, $n_{intermediate}$=2700, $n_{fast}$=3322). The best fits of the sum of two normal distributions to the data are shown as solid lines. The fluorescence peak values are located at $F(k)=F_0*2^k$, where k =0, 1, 2, 3. Cells under conditions of slow growth are assumed to typically have one origin at initiation of replication (see eg. Fig. S5A), with an associated fluorescence mass for each chromosome, $F_0$. Therefore $2^k$ is taken to be the number of origins, $n_{ori}$, and therefore the typical number of origins at initiation are 1, 2, 4 for slow, intermediate and fast growth conditions.

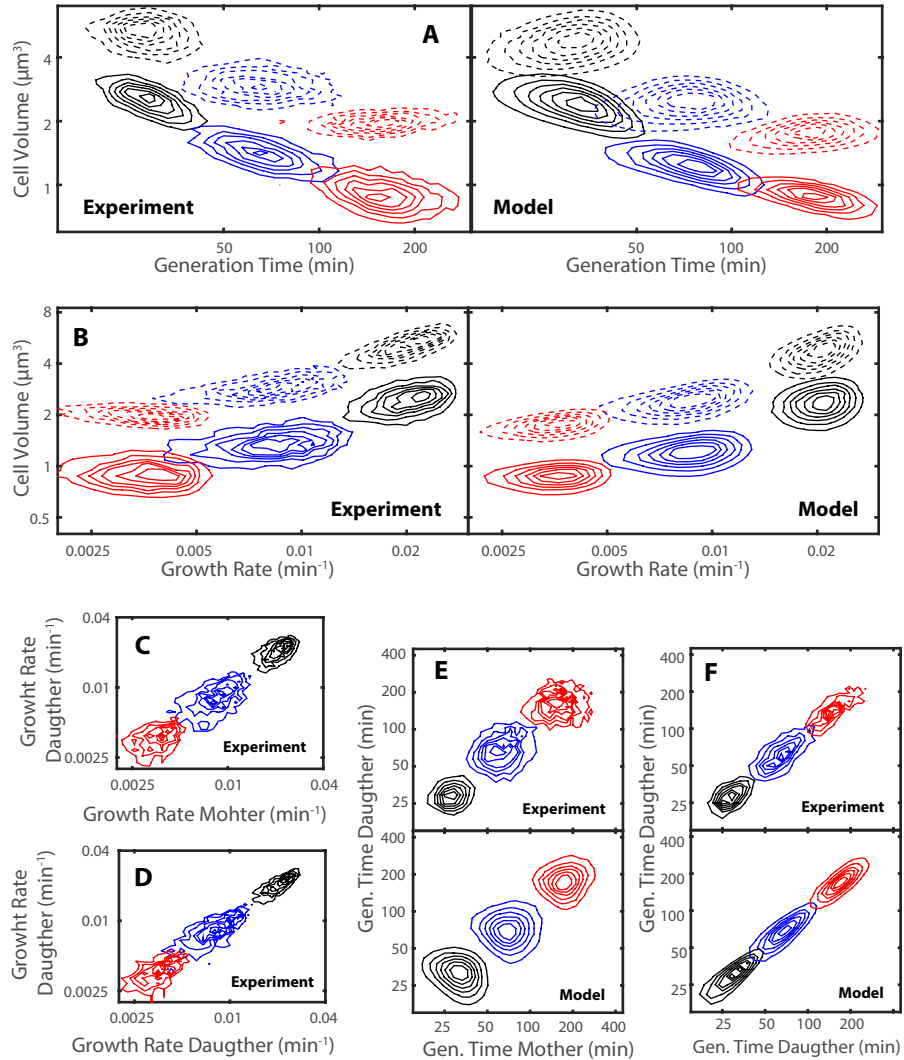

**Figure S7: Correlation between cell physiology parameters in experiments and model.** (**A**) Same as Fig. 4A in the main text, but with logarithmic axes. (**B**) Correlation between cell sizes and growth rate. (**C**) Correlations in generation times for mothers and daughters. The correlation coefficient, r=0.47, r=0.50, r=0.36 for the fast, intermediate and slow growth conditions respectively. (**D**) Correlations in generation times for sister cells. The correlation coefficient, r=0.69, r=0.69, r=0.56 for the fast, intermediate and slow growth conditions respectively. (**E**) Correlations between generation times in mother and daughters. (**F**) Correlations between generation times in the two daughters.

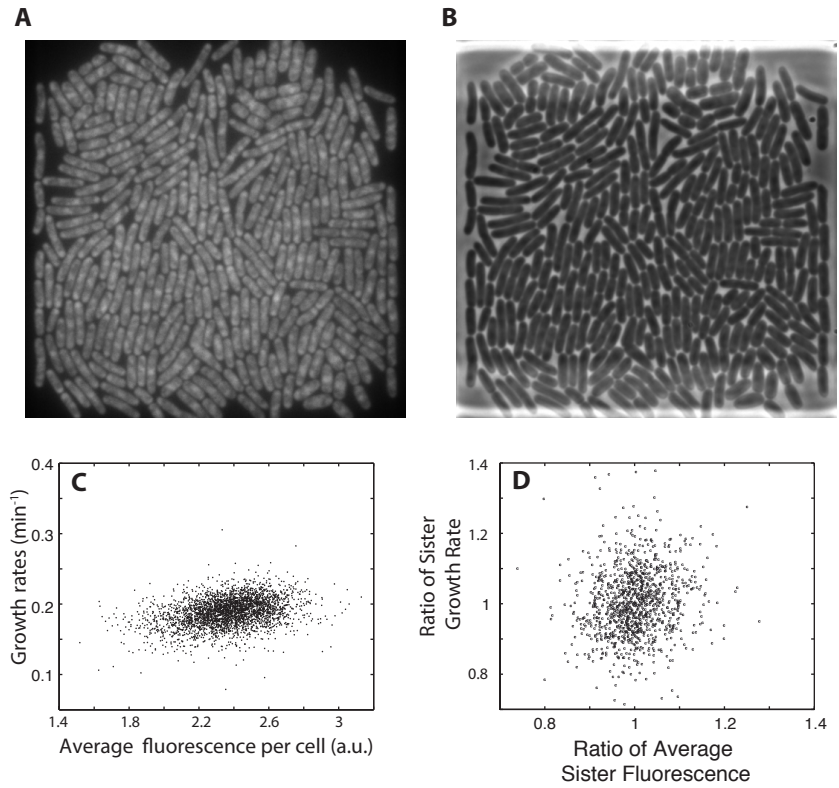

**Figure S8: Growth rate dependence on ribosomal content and partitioning at birth.** (**A**) Example of S2-Venus fluorescence. (**B**) Example of the corresponding phase contrast image. (**C**) Joint observations of growth rate and S2-Venus fluorescence at birth. The corresponding correlation coefficient is r=0.330, (n=1090). (**D**) Joint mapping of observations of the ratio of growth rate between sisters and the ratio of S2-Venus fluorescence at birth. The corresponding correlation coefficient, r=0.140 (n=1090).

```
begin
    t ← 0, V ← 1, n_ori ← 1, t_I ← 0, t_D ← ∞, t_T ← ∞,
    I ← 1;
    while t < t_max do
        μ ← random from Norm(⟨μ⟩, σ_μ) ensuring μ > 0;
        t_I ← t + (1/μ) log(I/V);
        repeat
            update both the stored division times and t_D (if not ∞) to
            the new growth rate accounting for the fraction of τ_D used
            in previous generation:
            t_D ← t + τ_D(μ) (t_D - t)/(t_D - t_T^S)
            where t_T^S is the termination time associated with each
            divison.
            if (t_I < t_D and t_I < t_T) or n_ori = 1 then
                t ← t_I;
                store t + τ_C in sorted list;
                t_T ← smallest value of stored termination times;
                V ← I;
                n_ori ← 2n_ori;
                V_I ← random from Norm(⟨V_I⟩, σ_{V_I}) ensuring V_I > 0;
                I ← V_I n_ori;
                t_I ← t + (1/μ) log(I/V);
            else if t_T < t_D then
                V ← V exp(μ(t_T - t));
                t ← t_T;
                store t + τ_D(μ) in sorted list;
                t_D ← smallest value of stored division times;
                remove t_T from the list of stored termination times;
                if no stored termination times then
                    t_T ← ∞
                else
                    t_T ← smallest value of stored termination times;
                end
            else
                V ← V exp(μ(t_D - t));
                t ← t_D;
                remove t_D from the list of stored division times;
                if no stored division times then
                    t_D ← ∞
                else
                    t_D ← smallest value of stored division times;
                end
            end
        until until divided;
        V ← V/2;
        n_ori ← n_ori/2;
        I ← I/2;
    end
end
```

**Variables and Functions:**

$\langle V_I \rangle$ : Average initiaition volume per number of origins.
$\sigma_{V_I}$ : Std. of initaition volume per nr. of origins.
$\tau_C$ : C-period.
$\tau_D(\mu)$ : Function returning growth-rate dependent D-periods (Fig 3F, main text)
$\langle \mu \rangle$ : Average growth-rate.
$\sigma_\mu$ : Growth-rate standard deviation.
$t_{max}$ : Max. simulation time.
$n_{ori}$ : Number of origins.
$t_I$ : Initiation time.
$\mu$ : Growth rate.
$t_T$ : Termination time.
$t_D$ : Division time.
$I$ : Initiation volume.
$t$ : Simulation time.
$V$ : Cell volume.

**Figure S9: Simulation algorithm.** The algorithm shows the case where there are no correlations between mother and daughter and in-between daughters have been included. The algorithm is easily extended account for mother-daughter and daughter-daughter correlations were the growth-rates were sampled from the correct multidimensional distribution. For the correlation between sisters one the algorithm was extended to follow both daughters, but only generating progeny from one. Averages and standard deviations as estimated in Fig. 2E of the main text: <μ>=0.021 min$^{-1}$, <μ>=0.009 min$^{-1}$, <μ>=0.0037 min$^{-1}$ for the fast, intermediate and slow growth conditions respectively. σ$_μ$=0.0038, σ$_μ$=0.0025, σ$_μ$=0.00089 for the fast, intermediate and slow growth conditions (Fig. 2E, main text). The mother-daughter and daughter-daughter correlations in growth-rates were as Figure S7. <$V_I$>=0.9 μm$^3$ (Fig 3, main text) and σ$_{VI}$=0.075•0.9 μm$^3$. C=50min, 70min and 120min for the fast, intermediate and slow growth conditions (Table S1).

# Supplementary Tables:

**Table S1: Parameters estimated from experimental observations**

| Condition | $\mu$ (min$^{-1}$) | $v_I$ (µm$^3$) | $v_I/n_{ori}$ (µm$^3$) | $v_T$ (µm$^3$) | $v_D$ (µm$^3$) | C (min) | D (min) | C+D (min) |
|---|---|---|---|---|---|---|---|---|
| Slow | 0.00371 ± 0.00003 | 0.90 | 0.9 | 1.42 | 1.80 | 123 | 62 | 185 |
| Slow R$_0$ | 0.00254 ± 0.00020 | 0.97 | 0.97 | 1.48 | 1.86 | 163 | 91 | 254 |
| Slow R$_1$ | 0.00333 ± 0.00013 | 0.93 | 0.93 | 1.43 | 1.80 | 130 | 68 | 198 |
| Slow R$_2$ | 0.00412 ± 0.00013 | 0.89 | 0.89 | 1.39 | 1.75 | 106 | 56 | 162 |
| Slow R$_3$ | 0.00486 ± 0.00019 | 0.83 | 0.83 | 1.39 | 1.77 | 106 | 50 | 156 |
| Intermediate | 0.00900 ± 0.00003 | 2.00 | 1.0 | 1.83 | 2.97 | 67 | 54 | 121 |
| Intermediate R$_0$ | 0.00579 ± 0.00048 | 2.05 | 1.03 | 1.63 | 2.70 | 80 | 87 | 167 |
| Intermediate R$_1$ | 0.00793 ± 0.00030 | 2.03 | 1.02 | 1.77 | 2.85 | 70 | 60 | 130 |
| Intermediate R$_2$ | 0.01009 ± 0.00031 | 1.95 | 0.98 | 1.90 | 3.08 | 66 | 48 | 114 |
| Intermediate R$_3$ | 0.01223 ± 0.00044 | 1.85 | 0.93 | 1.92 | 3.32 | 60 | 45 | 105 |
| Intermediate 37°C | 0.01484 ± 0.00013 | 2.00 | 1.0 | 2.40 | 3.15 | 59 | 18 | 78 |
| Fast | 0.02070 ± 0.00008 | 3.81 | 0.95 | 2.74 | 5.24 | 49 | 31 | 81 |

# Supplementary Online Material

## Materials and Methods

**Bacterial strains**
All bacterial strains studied were derived from *E. coli* MG1655 or its derivative BW25993 (*1*). Replisomes were investigated using strain JJC5350 (*2, 3*), carrying a genetic fusion of the replication factor *dnaQ* and the yellow fluorescent protein *ypet*, encoded in the native *dnaQ* locus. The construct was also transferred to strain BW25993 using a P1-phage.

The location of origins (see figure S5) was investigated using strain JE200, in which a genetic fusion of the transcription factor *malI* and the gene encoding the yellow fluorescent protein variant v*enus* was introduced in the origin-proximal *bgLG* locus in strain BW25993 using the lambda-red protocol (*1*). The construct contains two tandem operator sites, *malO*, to which MalI-Venus binds tightly. Further, the native *malI* gene as well as the native *malO* sites were deleted using the lambda-red method. This minimal construct was selected to avoid the risks associated to having a large number of operator-transcription factor complexes present in the cell (*4*).

Precision in the determination of division timing (see under *Accuracy and precision of the phase contrast division classifier* and figure S2) was investigated in strain EC442 (*5*), containing a genetic fusion of the division factor, *ftsQ,* and a green fluorescent protein, *gfp,* is introduced in MG1655. The construct is regulated by the inducible lactose promoter.

Accuracy in determining indual growth rates (see under *Determination of individual growth rate*) was investigated in strain JE201, in which a gene encoding a red fluorescence protein, *tRFP*, regulated by the constitutive ribosomal RNA promoter *P2rrnB* was introduced at the *intC* locus using the lambda red method in BW25993.

The dependence of the growth rate on ribosome content (see under *The dependence of growth rate and ribosome content and partitioning at birth*) was studied in strain JE202, in which gene *rpsB* was genetically fused to yellow fluorescence protein, *Venus*, and gene *rplI* to a red fluorescent protein, *mCherry*, using the lamda-red protocol. In both cases the constructs replaced the native genes. The *rpsB* and *rplI* genes expresses the S2 and L9 proteins, which associates to the small and large subunit of the ribosome respectively.

**Growth conditions**
Cells were grown in M9 minimal medium, unless stated otherwise. Fast growth was with 0.4% glucose supplemented with RPMI 1640 amino acids [R7131, Sigma-Aldrich] at 37°C. For intermediate growth, cells were grown with 0.4% succinate supplemented with RPMI 1640 amino acids [R7131, Sigma-Aldrich] at 30°C unless stated otherwise. During slow growth conditions, cells were grown with 0.4% acetate at 37°C. Strain EC442 was grown under fast conditions with 5 μM IPTG present in the medium to induce the expression of FtsQ-GFP molecules. Strain JE201 was grown in LB at 37°C. Strain JE202 was studied under fast growth conditions. All media used in the microfluidic experiments contained a surfactant, Pluronic F108 (CAS 9003-11-6, Sigma-Aldrich) to a final concentration of 0.85 gL$^{-1}$.

**Microfluidic sample management**
The preparation and operation of the microfluidic devices used in experiments with strains JJC5350, JE200, EC442, JE201 and JE202 was performed as described in (*6*). For experiments under slow and intermediate growth conditions, the trap depth used was 800 nm. For all other microfluidic experiments 900 nm was used.

**Microscopy & Imaging conditions**
All microscopy experiments were performed using an inverted microscope (NIKON TI-E) with an oil immersion objective (100X APO tirf 1.49 N.A.). For phase contrast imaging, external phase contrast was used. Cells were imaged in phase contrast using either CFW-1312M (Scion Corporation) or a DMK 23U274 (The imaging source Corporation) camera and the lamp of the microscope as illumination source. Phase contrast images were acquired with 125 ms exposure and at a rate of 0.033Hz unless stated otherwise. For fluorescence imaging of strain JJC5350 and JE200 a laser emitting at 514 nm (Coherent Genesis CX STM) was used to illuminate the sample. For strain JJC5350 DnaQ-YPet molecules in the sample were exposed for 1000 ms at an illumination power of 3.0 $Wcm^{-2}$ per frame. For strain JE200, bound MalI-Venus molecules were exposed for 4000 ms at an illumination power of 0.75 $Wcm^{-2}$. For both strains, cells were imaged for fluorescence at a frequency of 0.0056 Hz. The acquisition of each fluorescence frame was accompanied by the acquisition of a brightfield image using the configuration of the fluorescence channel with the exception that the microscope lamp was used as illumination source. Fluorescence and brightfield images were recorded on an Andor Ixon EM+ camera, unless stated otherwise. This camera was equipped with an additional 2xmagnifying lens, so that the final magnification of the sample was 200x. For strain EC442, FtsQ-GFP molecules in the sample were exposed using a laser emitting at 488 nm (Cobolt MLD) for 100 ms during at a frequency of 0.017 Hz. The microscope was controlled using micro-manager (*8*)(*8*) and automated acquisitions were performed using customized software, RITacquire 2.0. All time-lapsed acquisitions were performed in parallel at 7 trap regions simultaneously, one of which was not exposed to laser illumination unless stated otherwise. The duration of the acquisition varied from 2-24 hours depending on growth conditions. The temperature of the microfluidic device was maintained using cage incubator (OKO-lab) encapsulating the stage. For imaging of strain JE201, fluorescence imaging the sample was exposed using a laser emitting at 561 nm (genesis MX, Coherent) for 300 ms per frame at a frequency of 0.017 Hz. One trap was monitored in the experiment. For strain JE202, phase contrast images were acquired at a frequency of 0.1Hz. For fluorescence imaging a laser emitting at 514 nm (Fandango 150, Cobolt) was used to illuminate S2-Venus molecules in the sample at a power of 18.5 $Wcm^{-2}$, for 30 ms per frame. L9-mCherry molecules in the sample were exposed using a laser emitting at 580 nm (F-04306-03, MBP Com. inc) at power of 61.5 $Wcm^{-2}$ for 200 ms per frame. Fluorescence and brightfield images were acquired using and Andor Ultra at frequency of 0.056 Hz. Three traps were monitored in parallel. For the microfluidic run-out experiments, DNA-fluorescence was measured at 250xmagnification and using a lamp (Lumencor) as illumination source at 405 nm.

**Cell segmentation, tracking and detection of single molecules**
A custom automated analysis pipeline written in MATLAB 2015a was used to analyze the time-lapsed microscopy data. The edges of the trap in each series were identified in each phase contrast and each brightfield image and were used to superimpose information from each camera into a common coordinate system spanned by the trap. Phase contrast images were cropped to remove areas outside the traps as well as rescaled in intensity to reduce fluctuations. Cells in each phase contrast image were segmented using the method described in (*9*). A custom active contour model based on (*10*) was developed and a contour was computed for each segmented object. Cells were then tracked between frames using the method described in (*11*). The output was filtered so that a cell cycle must have must have a parent, have two children and that the maximum displacement of that cell between any consecutive frames was less than the cellwidth. An additional growth condition dependent filter criteria was added so that only cells with a cycle time greater than 50, 30 and 10 minutes were retained for slow, intermediate and fast growth respectively. Determination of length, areas, volumes and widths were based on the contour model as in (*10*). Replisomes and origins were detected in the raw fluorescence images using the method described in (*12*). The coordinates were set in a cellular reference frame using the contour model derived from the phase contrast image taken in conjunction with the fluorescence image.

**Microfluidic replication run-out experiments**
At the beginning of each experiment, BW25993 cells were grown in the microfluidic device as before. Using the technique described in (*12*), the medium was rapidly switched (~2 s) to medium containing 300 µg/mL Rifampicin (CAS 13292-46-1, Sigma-Aldrich) and 30 µg /mL Cephalexin (CAS 23325-78-2, Sigma-Aldrich). Cells were then incubated for 2-3 hours to allow DNA replication to complete and the effects of the exchange were monitored by time-lapsed phase contrast microscopy. No visible pressure perturbation was observed at the instant medium was exchanged. Cells were fixed by manually disconnecting the tubing supplying the medium to the device and exchanging this to instead supply a solution of 3.6% formaldehyde in water to the growth chamber. Cells were subjected to fixation for 30-60 minutes at the temperature associated with each growth condition. The same procedure was used to exchange the fixation solution to a solution of 1.5µg/ml Hoechst 34580 (Molecular Probes) in PBS (pH=7.3) to stain the DNA content of the cell. The fixed cells were stained for 2-3 hours before measuring the fluorescence resulting from the DNA stain. Cells were segmented from raw fluorescence images and the corresponding fluorescence values were integrated.

We note that when Rifampicin and Cephalexin are introduced in the run-out experiment, a large fraction of cells lyse. Further, when performing the experiment for fast growth, exchanging medium to Cephalexin (10 µg/ml), we noted that cells continue to divide at least once after the exchange.

**Replication run-out experiments on flask cultures**
For each strain a culture of LB was inoculated from glycerol stocks kept at -80°C and grown at 37°C and shaking until stationary state was reached. 250 µl of each such culture was then used to inoculate 50 ml of medium in a 100 ml E-flask corresponding to slow, intermediate and fast growth conditions. Cultures were incubated at the corresponding temperatures with shaking until the optical density at 600 nm had reached 0.1-0.3. This took, 12.5, 5, and 2 hours for slow, intermediate and fast growth respectively. Rifampicin and Cephalexin were added to each culture to final concentrations of 300 µg/ml and 30 µg/ml and the cultures were then incubated for an additional 2 hours in the case of intermediate and fast growth and 3.5 hours in the case of slow growth. Cells were harvested by centrifugation at 5000g for 3 minutes and re-suspended in 1 ml 3.6% formaldehyde and incubated at room temperature for 30-60 minutes. Cells were then centrifuged 1-2 min at 5000g and re-suspended in 1 ml PBS (pH 7.3) with 1.5 µg/ml Hoechst 34580 (Molecular Probes) and incubated over-night at 4°C while covered so as not to bleach prior to measurement. DNA content was measured using a BDFACS Aria IIu (BD Biosciences) cytometer using a nozzle of width 70 µm. Samples were diluted in PBS (pH 7.3) so that averaged event rates were within 10000-30000s$^{-1}$ and measured using standard DAPI illumination and detection configurations.

**Validation of the phase contrast division classifier**
At each fluorescence frame, an evenly spaced grid of points was spanned to cover each cell and the fluorescence intensity values at these points were computed by interpolation (see figure S2A top). The values along the short axis of the grid were averaged at each point of the long axis, yielding the fluorescence intensity over the length of each cell from pole to pole. The z-score was computed for this long axis intensity trace, which was then combined with those obtained at different time points for the same cell (see figure S2A bottom). FtsQ-GFP was expected to accumulate at the septation point. This peak in the density (see figure S2A bottom) would typically vanish, often between two consecutive fluorescence frames (1 min). This was taken as the division time for the cell. The location of the FtsQ-GFP band was identified by selecting the point along the main axis with the greatest power for low frequencies over time in the Fourier spectrum. The division time was determined as the time point at which the most rapid decrease in FtsQ-GFP signal occurred. To continue the analysis beyond the division time determined by the phase contrast classifier contours of the two daughters were merged into one contour, which was then analyzed as described above. The maximum time-limit for merging daughters was set 10 minutes post phase contrast division.

**Characterization of growth**
The growth in a flask culture was studied for strain BW25993, JJC5350, JE200 and BW25993 dnaQ-YPet, by repeatedly sampling liquid cultures of the medium corresponding to fast growth conditions, incubated with shaking (see figure S1A). The optical density at 600 nm, $OD_{600}$, was measured for each sample using a spectrophotometer (BioRad smartspec plus). The maximum growth rate was determined linear regression to the natural logarithm of the measured values as a function of time.

The growth of each strain was also studied by time-lapsed measurements of $OD_{600}$, in a plate reader (TECAN inf M200). For each strain 1000 µl cultures corresponding to slow, intermediary and fast growth were inoculated with 1 µl overnight LB culture. Each sample was studied in triplicates and each consisting of 200 µl culture. The samples were incubated at 37°C for 48 hours, while repeatedly (0.2 $min^{-1}$) first shaking the culture, then measuring $OD_{600}$. The maximal growth rate, µ, was determined for fast and intermediate growth conditions for each strain and replicate according to the Gompertz growth law (*13*),

$y = y_f \cdot exp(-exp((\mu \cdot exp(1)/y_f) \cdot (t_L - t) + 1))$

where t is the time from the start of the experiment, y is the natural logarithm of $OD_{600}(t)/OD_{600}(t=0)$, $y_f$ is the final value of y and $t_L$ is the lag time. The measurements for slow growth do not conform to the Gompertz form, possibly as a consequence of evaporation of medium over 48 hours.

**Determination of individual growth rate**
The growth rate of each cell was computed by fitting the following expression to the data

$L = L_B \cdot exp((t - t_B) \cdot \mu)$

Here, L is the length of the cell at time t, $L_B$ and $t_B$ are the length at birth and the time of birth respectively. The accuracy in determining individual growth rates was studied using strain JE201. Cells were segmented both from phase contrast and fluorescence images. A backbone of each cell was calculated as the best fit of a second order polynomial and the length of each cell was determined as the integrated distance from pole to pole along the backbone.

**Determination of initiation and termination sizes**
The initiation size, $V_I$, was estimated for each growth condition from the average number of observed replisomes, y, per cell as a function of the of the corresponding cell volume. In the cases of slow growth the data was fitted to the following expression

$y = y0 \cdot exp(-½ \cdot (V - V_p)/s_p)^2)$

where $V_p$ is the volume at which the replisome density peaks and $s_p$ is the width of the peak. The initiation volume $V_I$ and the termination volume $V_T$ were determined as

$V_I = V_p - (2 \cdot s_p^{2 \cdot ln(2)})^{-1/2}$
$V_T = V_p + (2 \cdot s_p^{2 \cdot ln(2)})^{-1/2}$

During intermediate (30°C and 37°C) and fast growth the replication cycles overlap, initiation volume was instead determined from the best fit of

$y = y0 + y1 \cdot (1 + erf((V - V_I)/(s_I \cdot 2^{-1/2})))$

where, erf is the error function and $s_I$ is the spread of initiation. The termination volume, $V_T$, were determined from the average number of replisomes per cell observed within ± 0.5 μm of the midsection. The data was fitted to

$$y = y_0 + y_1 \cdot (1.5 - 0.5 \cdot \text{erf}((V-V_T)/(s \cdot 2^{-1/2})))$$

For slow and intermediate growth conditions the data was stratified into categories based on growth rate. Four categories for each growth condition were established as

$R_0: m_\mu - 2 \cdot \sigma_\mu \leq \mu < m_\mu - 1 \cdot \sigma_\mu$
$R_1: m_\mu - 1 \cdot \sigma_\mu \leq \mu < m_\mu$
$R_2: m_\mu \leq \mu < m_\mu + 1 \cdot \sigma_\mu$
$R_3: m_\mu + 1 \cdot \sigma_\mu \leq \mu < m_\mu + 2 \cdot \sigma_\mu$

where μ is the growth rate, $m_\mu$ and $\sigma_\mu$ are the average growth rate standard deviation of the entire population respectively.

**Determining the C + D period**

The C period was determined as

$$C = \mu^{-1} \cdot \ln(n_{ori} \cdot (V_T/V_I))$$

where $n_{ori}$ is the typical number of origins at initiation for each growth condition as determined by the microfluidic run-out experiments. For intermediate growth conditions at 37°C were assumed to have the same $n_{ori}$ as at 30°C. The C+D period was determined as

$$C+D = \mu^{-1} \cdot \ln(n_{ori} \cdot (V_D/V_I))$$

where $V_D$ is the volume at division for each growth condition. For slow growth the contour model of each cell misrepresents the volume prior to division as the complement to the septum is not accounted for. The volume of the septum complement is estimated as

$$V_{sc} = (2/3) \cdot \pi \cdot r^3$$

where r is the average cell radius and is subtracted from the division volume derived from the contour model. For stratified data the median rather than the mean was used to determine μ and $V_D$ for each category. For intermediate growth an additional data set was included where the temperature was 37°C instead of 30°C. This set was analyzed in the same way as the data from the intermediate growth conditions. Fitting of the data to the functional form $C+D = \alpha \cdot \mu^{-\beta} + \gamma$. was done using the results of the stratified categories for slow and intermediate (30°C) growth instead of the average of the entire population. During fitting, each stratified category was assigned a weight of 0.25, while the values for intermediate growth at 37°C and those for fast growth were assigned a weight of 1.

**The dependence of growth rate on ribosome content and partitioning at birth**
Imaged cells of strain JE202 were segmented, tracked and their growth rate was determined as described earlier. The total S2-Venus fluorescence corresponding to each cell in the earliest available fluorescence image was divided by the area of the cell to form the average S2-Venus fluorescence at birth.

**Supplementary Note: An ultrasensitive mechanism for triggering replication initiation at a fixed volume to chromosome ratio.**

The regulatory problem of synchronously initiating replication in responses to minor changes is the chromosome to volume ratio is very challenging. It will require a striking regulatory principle to achieve extremely sensitive, yet robust, initiation response at all growth rates(*14*). Such a mechanism has not previously been described, and for this reason we will here outline a plausible regulatory principle that fulfils these requirements.

Below we will introduce the major players that are important for the initiation mechanism. For an in depth review of the molecular biology of replication control, please see (*7, 15*). The main activator of replication initiation is the protein DnaA that binds at multiple specific sites at OriC and also at many specific sites at the rest of the chromosome, especially in the *datA* region close to OriC. DnaA binds tightly to ATP or ADP, but only the ATP bound form can initiate replication by forming a multimer at OriC. For this reason DnaA-ATP has been implicated to be the main regulator of replication initiation and most models of replication initiation have been focused on titrating the DnaA-ATP concentration to a critical initiation concentration. After initiation, the hemi-methylated GATC sequences are sequestered by SecA, which prevents immediate reinitiation at the same OriC and thus allows each OriCs to fire only once (*16, 17*). Before the sequestration period is over the initiation potential also drops since free ATP-DnaA binds at *datA* and other sites at the newly replicated chromosome. During the passage of the replication fork the previously bound ATP-DnaA is converted to ADP-DnaA by the RIDA (regulated inactivation of DnaA) mechanism. Thus the questions of how multiple OriC can initiate close in time without reinitiating on the same OriC is largely resolved, the challenging question is what brings the initiation potential back up at the correct growth rate dependant time?

How sensitive does the initiation potential have to respond to the chromosome volume ratio in order to produce stable replication-division cycles? OriC is sequestered and inert for new initiation approximately one third of the cell cycle (*16*) and to achieve synchrony it is thus critical that each OriC initiate during this, and only this time. Wild-type cells fails in synchronous initiation only in a few percent of the cases(*18*). To achieve less than 1% asynchronous initiation under conditions with four 4 origins, it means that for 4 independent events has to occur within one third of the cell cycle with over 99% probability. This requires that each OriC must initiate with a rate of more than 30 per generation during this time. Furthermore, the initiation rate has to be less than 0.01 per generation per OriC during the rest of the cell cycle so that non of the OriC fires at the wrong time, *i.e.* the initiation rate increase more than 3000 fold in response to a less than 50% change in the chromosome to volume ratio (i.e the volume change occurring in one third of the cell cycle). In total the required sensitivity of regulatory system needs to be in the order or 6000 to accomplish this task. Sensitivity is here defined as sensitivity amplification factor(*19*), i.e. the relative change in the initiation rate compared to a relative change in the chromosome to volume ratio.

A sensitivity amplification of 6000 is non-trivial to achieve. The concerted binding and polymer formation of DnaA-ATP at OriC may form the basis for a cooperative, or multistep, control scheme where DnaA-ATP polymerization can be interrupted by a single DnaA-ADP incorporation(*20*). However, given that only about 10 DnaA molecules can bind at OriC, is seems far stretched to imaging getting much higher sensitivity amplification then 10 in initiation rate to DnaA-ATP concentration or initiation rate to DnaA-ATP/ADP ratio, A factor of a 1000 thus remains to be explained in how the chromosome to volume ratio sets the concentration of active DnaA.

Neither of the conventional schemes for expression of an inhibitor or an activator of replication can achieve the required level of sensitivity for DnaA-ATP response to chromosome volume ratio. For example, expression of a repressor of initiation in proportion the chromosome copy number that is diluted by volume growth(*21*), would give a sensitivity of 1. Similarly, relying on that conversion of DnaA-ATP into DnaA-ADP by the RIDA mechanism and that the initiation potential goes back when the DnaA-ATP concentration recovers by growth (*22*) also has a sensitivity in the order of 1. Thus, it is reasonable to believe that the *E. coli* cell is using another mechanism.

One way, possibly the only way, to reach sufficiently high sensitivity regulation is to use a zero-order switch mechanism (*23*) based on the cycling of DnaA-ATP and DnaA-ADP. If the conversion flux of DnaA-ATP to DnaA-ADP is proportional to the number of chromosomes and the DnaA-ADP to DnaA-ATP flux is proportional to the volume (see cartoon below), the sensitivity amplification that can be achieved in the DnaA-ADP concentration in response to a change in the chromosome to volume ratio is very high. In fact, is approximately equal to the total DnaA concentration divided by the $K_M$-value in the modifying reactions (*24, 25*). By making the actual initiation reaction respond to the ratio of DnaA-ATP to DnaA-ADT further sensitivity is gained.

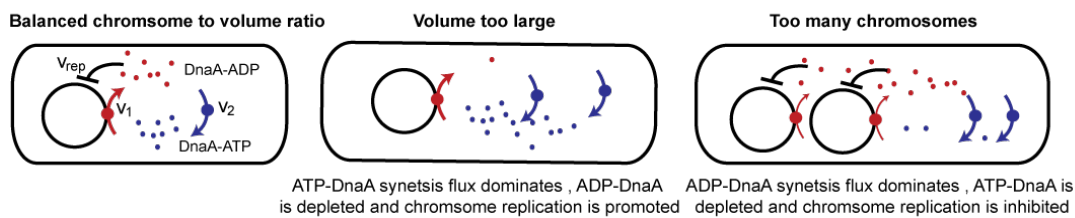

For the conversion of DnaA-ATP to DnaA-ADP exactly such an enzymatic process has by Katayama (*26*) been described in the DDAH (datA-dependent DnaA-ATP hydrolysis) mechanism, where IHF bound to the chromosomal *datA* locus converts DnaA-ATP to DnaA-ADP at a rate that is very insensitive to the DnaA-ATP concentration. When this mechanism is combined with a reaction for converting DnaA-ADP to back to DnaA-ATP, which is mediated by an enzyme of relatively fixed concentration that is easily saturated by DnaA-ADP, it makes for a zero–order ultrasensitive switch. The hypothetical DnaA-ADP to DnaA-ATP conversion could, for example, depend on the IHF dependant DARS mechanism (*27*) but it is critical that the conversion rate is proportional to the volume of the cell not the number of DARS sites.

Since the cells is growing and thus not in steady state with respect to the chromosome to volume ratios, analytical solutions are hard to come by. Instead we test the suggested regulatory scheme using stochastic simulations (figure below) Here, the exact initiation timing is probabilistic in response to the time-variable initiation rate, but the different growth rates are set fixed in order to evaluate how accurately the initiation volume is given by the mechanism.

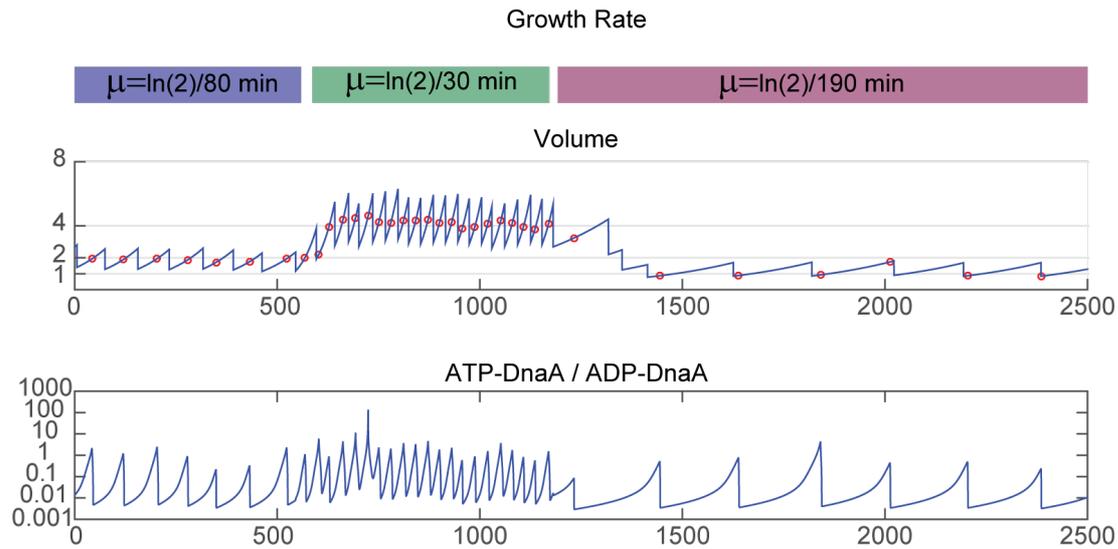

*Modelling assumptions*

Let replication initiate at a rate $v_{rep}$ based on to the DnaA-ATP to DnaA-ADP ratio:

$$v_{rep} = k_{ini} \text{ DnaA-ATP}^2/(\text{DnaA-ADP}+\text{DnaA-ATP})^2.$$

We use $k_{ini}=0.01s^{-1}$ per OriC, implying that that the probability of initiation rate per OriC is 1% in a one second time-period when the DnaA-ATP >> DnaA-ADP. The exact value of $k_{ini}$ is not important as long as the initiation potential drops within the sequestration time, which is set to 1000 seconds from replication initiation. A Hill coefficient of 2 can, for example, be achieved by a multi-step initiation scheme at OriC that can be interrupted in 2 of places by the entry of DnaA-ADP instead of DnaA-ATP.

DnaA-ATP is converted to into DnaA-ADP by the enzymatic DDAH mechanism located at *datA* with a rate

$$v_1 = k_{ATP} N [\text{DnaA-ATP}]/(K_m + [\text{DnaA-ATP}]),$$

The rate is proportional to the number of *datA* sites, N, and saturates at $k_{ATP}$ (here $k_{ATP}=16$nM/s) when [DnaA-ATP]>>$K_M$ (here $K_M=5$nM). Similarly for the ADP to ATP cycling:

$$v_2 = k_{ADP} [\text{DnaA-ADP}]/(K_M + [\text{DnaA-ADP}])$$

which saturates at rate $k_{ADP}$ (Here $k_{ADP}=10$nM/s) when [DnaA-ADP]>>$K_M$ (Here $K_M=5$nM). The exact values of $k_{ATP}$, $k_{ADP}$ are not important as long as they allow sufficient cycling rate (>3nM/s). Their ratio sets the average initiation volume, here ~1fL per *datA*.

The mechanism is exceedingly robust to changes in total DnaA concertation over the physiological range (300nM to 10µM). For the demonstration the growth rate is set to ln(2)/30min, ln(2)/80min, and ln(2)/190min and corresponding C+D-periods are set to 74, 113 and 181min respectively. In practice all relevant values works. The standard-deviation to mean ratio in the initiation volume is for these parameters ~0.05.

*Final note:* There will surely be backup systems that kick in under certain extreme conditions but we propose that a zero order switch mechanism is the main principle of how to get sufficient sensitivity. This model is based mainly on what is theoretically required to achieve the observed sensitivity, but it also reproduces the main experimental perturbations that has been done to the replication system, *i.e.* deletion of *datA* results in over initiation, insensitivity to additional copies of OriC either on mini chromosomes or on the chromosome(*28*), no effect of a couple of times over-expression of DnaA (*29*), no effect on initiation timing after removal of the major DnaA binding sites outside datA (*30*)